\documentclass[pdftex, twocolumn]{jpsj3}
\usepackage{txfonts}
\usepackage{graphicx}
\usepackage{amsmath}
\usepackage{amssymb}

\title{Magnetoelectric Effect Dependent on Electric Field Direction in a Pyroelectric Ferrimagnet CaBaCo$_4$O$_7$}

\author{Takumi Shirasaki$^1$, Masaaki Noda$^1$, Hinata Arai$^1$, Mitsuru Akaki$^2$, Haruhiko Kuroe$^1$, and Hideki Kuwahara$^1$\thanks{h-kuwaha@sophia.ac.jp}}
\inst{$^1$Physics Division, Sophia University, Tokyo 102-8554, Japan\\
$^2$Institute for Materials Research, Tohoku University, Sendai 980-8577, Japan}

\abst{
    This study investigates the dependence of the static magnetoelectric (ME) effect on the external field direction in the pyroelectric-ferrimagnet CaBaCo$_4$O$_7$, a topic that remains largely unexplored compared to dynamical nonreciprocal ME effects. 
    We measured the magnetization with respect to the inherent polarization and found that an external electric field stabilizes the ferrimagnetic phase when applied parallel to the polarization, and destabilizes it when antiparallel. 
    These results clearly demonstrate the electric-field-direction dependent control of the static ME effect, suggesting a new route to enhancing ME effects in pyroelectric-magnetic materials. 
}

\begin{document}
\maketitle
    Since the discovery of the gigantic magnetoelectric (ME) effect originating from the spin structure in TbMnO$_3$, multiferroic materials exhibiting a cross-correlation between the magnetic (electric) field and ferroelectricity (ferromagnetism) have become a major research topic in condensed matter physics\cite{Kimura_Nature2003,Fiebig_JPhysD2005,Tokura_RepProgPhys2014}. 
    In multiferroic materials where both time-reversal and space-inversion symmetries are simultaneously broken, a type of magnetoelectric-optical effect known as nonreciprocal directional dichroism (NDD) is observed. 
    NDD is a phenomenon where the absorption rate of the electromagnetic waves depends on the sign of the term $\boldsymbol{k}\cdot\boldsymbol{P}\times\boldsymbol{M}$, where $\boldsymbol{k}$ is the propagation vector of the electromagnetic wave, $\boldsymbol{P}$ is the electric polarization, and $\boldsymbol{M}$ is the magnetization\cite{Hornreich_PR1968,Kimura_APLM2023}. 
    This dependence allows for the control of the absorption rate by reversing external fields or the propagation direction of electromagnetic waves. 
    NDD has been observed in many multiferroic materials, such as GaFeO$_3$\cite{Kubota_PRL2004}, CuB$_2$O$_4$\cite{Saito_JPSJ2008}, and $X_2$Co$Y_2$O$_7$ ($X$ = Ca, Sr, $Y$ = Si, Ge)\cite{Kezsmarki_NatCommun2014, Akaki_SciAdo2025}. 
    Moreover, this phenomenon has attracted considerable attention because it can be applied to optical devices such as optical isolators. 

\begin{figure*}[ht]
        \centering
        \includegraphics[height=55mm]{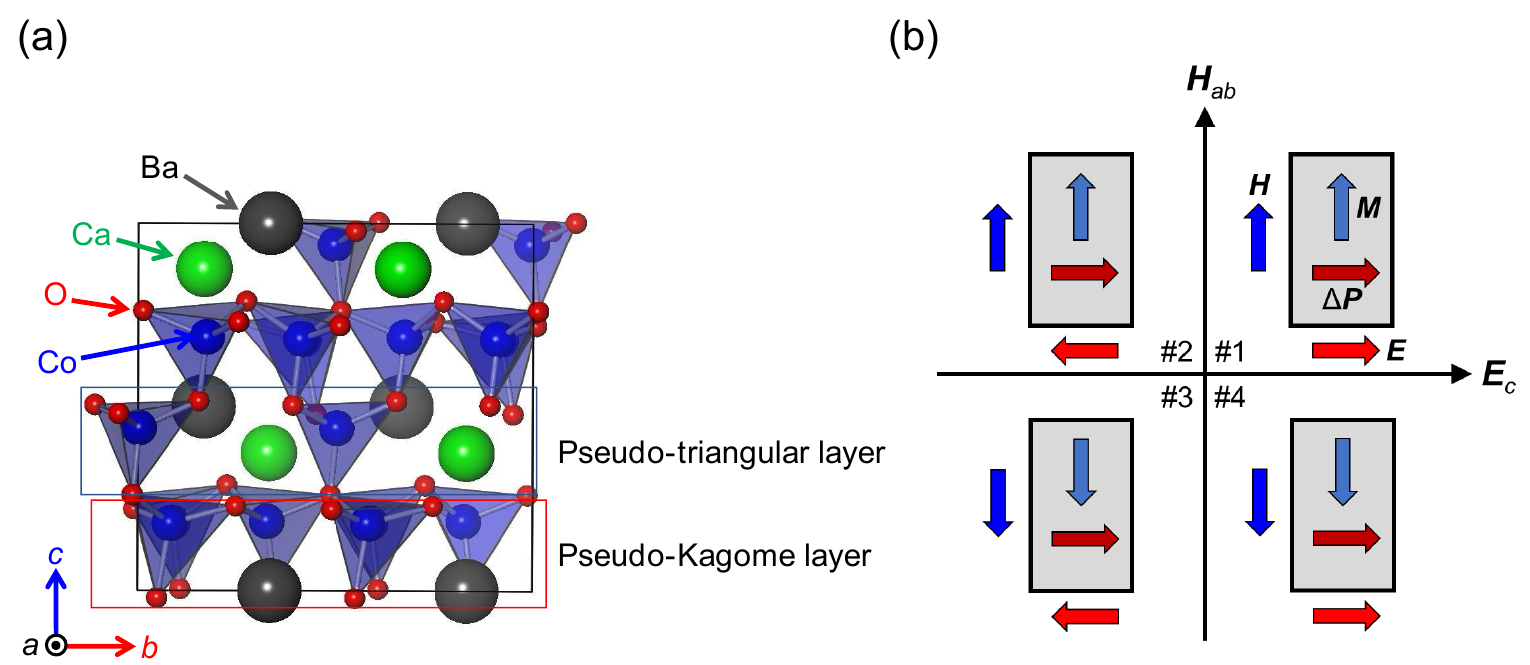}
        \caption{
        (Color online) (a) Schematic image of the CaBaCo$_4$O$_7$ crystal structure at room temperature (RT) visualized using the VESTA program\cite{Momma_JAC2011}. 
        The black rectangular box represents a unit cell of the \textit{Pbn}2$_1$ structure. 
        Pseudo-triangular and pseudo-Kagome layers are stacked along the \textit{c} axis to align the CoO$_4$ tetrahedra in one direction. 
        Therefore, the space-inversion symmetry is broken even at RT, and spontaneous electric polarization occurs along the \textit{c} axis. 
        (b) Schematic image of measurement arrangements. 
        Four measurement arrangements, labeled \#1 to \#4, can be considered based on different combinations of the directions of the external magnetic and electric fields ($\boldsymbol{H}_{ab}$ and $\boldsymbol{E}_c$, respectively). 
        The arrows for change in spontaneous electric polarization $\Delta \boldsymbol{P}$ represent the directions assumed to explain our experimental results consistently. 
        We assume a parallel configuration between $\boldsymbol{E}_c$ and $\Delta \boldsymbol{P}$ for arrangements \#1 and \#4, and an antiparallel one for \#2 and \#3. 
        On the other hand, $\boldsymbol{H}_{ab}$ and magnetization $\boldsymbol{M}$ are always parallel.
        }
        \label{CS}
    \end{figure*}

    NDD is a dynamical ME effect that depends on the application direction of external fields. 
    We focused on the influence of the application direction of external fields on the static ME effect. 
    Because of the ferroelectricity of many multiferroic materials, the direction of the external electric field forcibly reverses that of the induced electric polarization. 
    In pyroelectric materials, on the other hand, the direction of spontaneous electric polarization remains unchanged upon reversal of the external electric field, allowing for two distinct arrangements: one where the external electric field is parallel to the spontaneous electric polarization, and another where it is antiparallel. 
    In this study, we report that the magnetic properties of the pyroelectric-ferrimagnetic material CaBaCo$_4$O$_7$ can be controlled by the direction of the applied electric field, a behavior that can be consistently explained by assuming a fixed direction for the change in the spontaneous electric polarization $\Delta \boldsymbol{P}$, as schematically depicted in Fig.~\ref{CS}(b)\@.
  
    The crystal structure of the focused pyroelectric-ferrimagnetic material CaBaCo$_4$O$_7$ can be classified into the polar space group \textit{Pbn}2$_1$ at room temperature (RT), as shown in Fig.~\ref{CS}(a). 
    Co$^{2+}$ and Co$^{3+}$ spins form pseudo-triangular and pseudo-Kagome lattices, respectively, which exhibit strong geometrical frustration. 
    This material undergoes a ferrimagnetic phase transition at \textit{T$_{\rm{C}}$} = 63 K\cite{Iwamoto_JPC2012}. 
    Powder neutron diffraction measurements conducted on CaBaCo$_4$O$_7$ revealed that the magnetic easy and hard axes are the \textit{b} and \textit{c} axes, respectively\cite{Caignaert_PRB2010}. 
    A substantial change in spontaneous electric polarization along the \textit{c} axis was observed at \textit{T$_{\rm{C}}$} in CaBaCo$_4$O$_7$\cite{Iwamoto_JPC2012}. 
    This large ME effect can be attributed to the magnetostriction associated with the ferrimagnetic order, which triggers atomic displacement, and the resultant change in spontaneous electric polarization along the \textit{c} axis in the original polar crystal lattice\cite{Caignaert_PRB2013}. 
    
    Recent research has revealed that CaBaCo$_4$O$_7$ exhibits an antiferromagnetic order between \textit{T$_{\rm{C}}$} and \textit{T$_{\rm{N}}$} = 69 K, and a large ME effect is induced due to a phase transition from antiferromagnetic to ferrimagnetic by application of magnetic fields\cite{Omi_PRB2021}. 
    Furthermore, numerous elemental substitutions at the Ca and Co sites have been carried out, and it has been revealed that even a small amount of impurity substitution can significantly destabilize the ferrimagnetic ground state of CaBaCo$_4$O$_7$\cite{Sarkar_JMatChem2012,Seikh_PRB2012,Oda_PhysProc2015,Dhanasekhar_PRB2017,Gen_PRB2022}\@. 
    Additionally, CaBaCo$_4$O$_7$ breaks both time-reversal and space-inversion symmetries below \textit{T$_{\rm{C}}$}, leading to the emergence of NDD\@. 
    Indeed, NDD has been experimentally observed in the terahertz region\cite{Bordacs_PRB2015}. 
    CaBaCo$_4$O$_7$ is a pyroelectric material, not a ferroelectric one, and it exhibits a significant ME effect\cite{Johnson_PRB2014}.
    These properties make it a suitable material for investigating the influence of the application direction of external fields on the ME effect. 
    Although electric polarization is not reversed by reversing the electric field in such pyroelectric-magnetic materials, they will also open up new functionalities for spintronics devices if a significant dependence of electric field direction on the ME effect can be found. 

    The CaBaCo$_4$O$_7$ single crystal was synthesized using the following method. 
    The starting materials, CaCO$_3$, BaCO$_3$, and Co$_3$O$_4$ powders, were weighed and mixed with ethanol in a prescribed ratio, then pre-calcined for decarbonation at 900 $^{\circ}$C in air at ambient pressure for 12 h. 
    Next, the resultant powder was reground and pressed into a rod using hydrostatic pressure, and the rod was calcined at 1100 $^{\rm{o}}$C in an air atmosphere for 12 h and subsequently quenched to RT\@. 
    The single crystal was grown at a rate of 1 mm/h using the optical floating-zone method in air at a pressure of 3.5 atm. 
    Part of the obtained crystal was pulverized into powder, which was characterized by X-ray diffraction (XRD) using Cu-$K\alpha$ radiation (Bruker AXS, D8 Advance) at RT\@. 
    Rietveld analysis of the XRD patterns obtained from the pulverized crystals using the TOPAS program indicated that the grown crystal contained a single phase without any detectable impurity phases \cite{Coelho_JAC2018,Izumi_MSF2000}. 
    The crystal orientation was determined using the back-Laue method (Rigaku, Rint 2100).  
    
    To investigate the dependence of the ME effect on the application direction of external fields, we defined four measurement arrangements, \#1 to \#4 (see Fig.~\ref{CS}(b)). 
    A comparison of arrangements \#1 and \#2 can clarify the electric-field-direction dependence of the ME effect. 
    Meanwhile, comparing \#1 and \#4 allows for an investigation of how the ME effect changes with the sign reversal of $\boldsymbol{P}\times\boldsymbol{M}$. 
    The temperature and magnetic field dependences of the magnetization were measured using a physical property measurement system (Quantum Design, PPMS), which is a commercial apparatus. 
    Magnetization in electric and magnetic fields, which requires high accuracy, was measured using a SQUID magnetometer (Quantum Design, MPMS). 
    For all magnetization measurements presented hereafter, a common poling procedure was employed to align the magnetic domains. 
    The sample was cooled under field-cooling conditions with a magnetic field applied perpendicular to the $c$ axis and a poling electric field with strengths of $\pm$2 MV/m applied along the $c$ axis. 
    Due to the semiconductor-like electrical conductivity, the electric field was applied starting from 85 K, which is well above \textit{T$_{\rm{C}}$}, and where the sample possessed a sufficiently high resistivity.
    Because of the orthorhombic twinned structure of the CaBaCo$_4$O$_7$ single crystal, the \textit{a} and \textit{b} axes of the crystal could not be distinguished\cite{Omi_PRB2021, Arai_JPSC2022}. 
    Therefore, the magnetization and external magnetic field perpendicular to the \textit{c} axis are denoted by $\boldsymbol{M}_{ab}$ and $\boldsymbol{H}_{ab}$, respectively.

    \begin{figure}
        \centering
        \includegraphics[height=50mm]{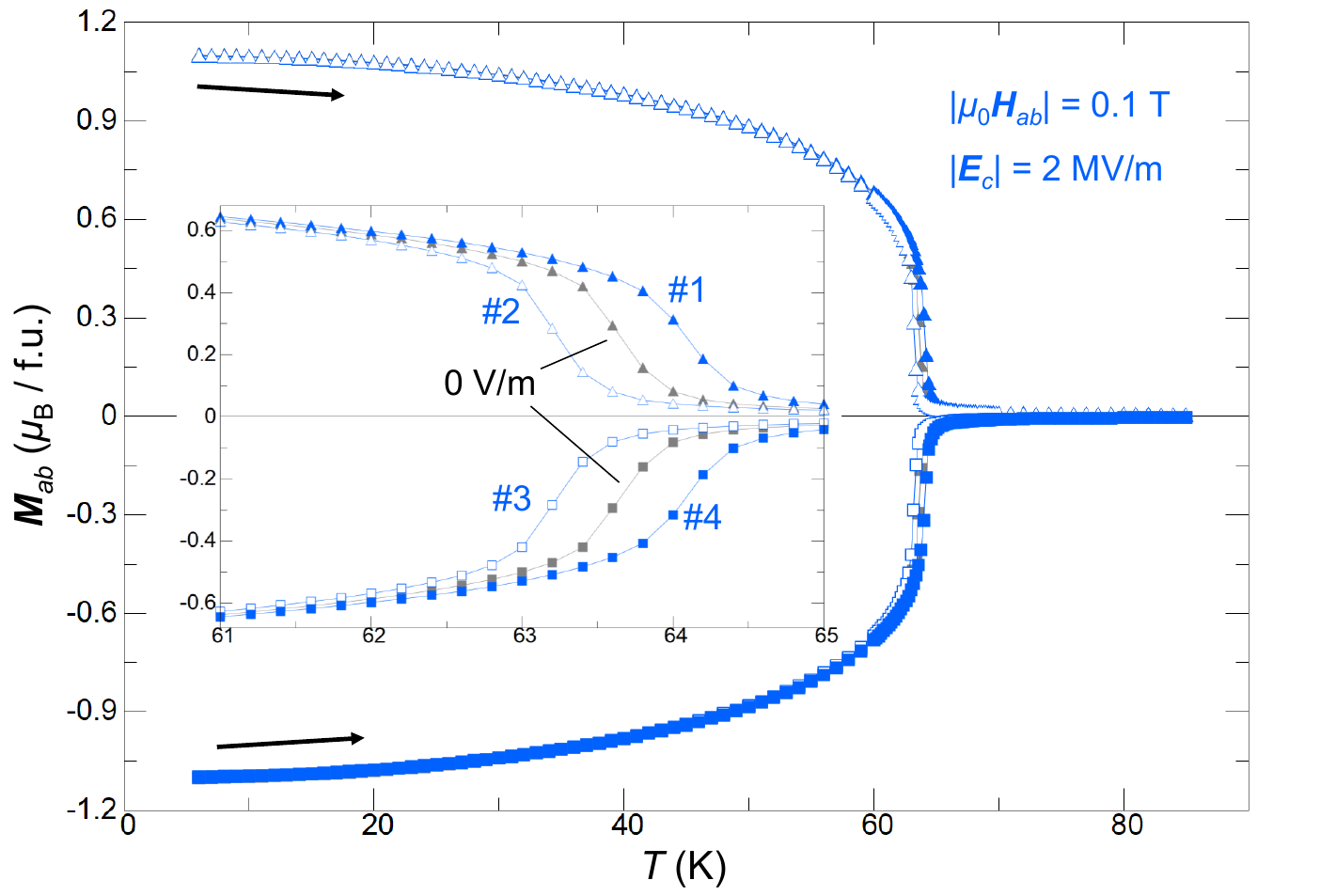}
        \caption{
        (Color online) Temperature dependence of magnetization in magnetic and electric fields. 
        The blue solid and open symbols correspond to the application of positive and negative external electric fields, respectively. 
        Furthermore, the triangular and square plots correspond to positive and negative directions of the applied external magnetic field, respectively, as shown in Fig.~\ref{CS}(b)\@. 
        The gray plots represent the measurements taken in zero electric field and magnetic fields of $\pm$ 0.1 T\@. 
        Inset is an expanded graph near $T_{\rm{C}}$.
        }
        \label{MT}
    \end{figure}
    Figure~\ref{MT} presents the temperature dependence of magnetization for arrangements \#1 to \#4, measured under combined electric and magnetic fields ($\boldsymbol{M}-T$ in $\boldsymbol{H}$ and $\boldsymbol{E}$). 
    The sample was cooled under each electric and magnetic field, and measurements were recorded during temperature-warming scans without removing external fields. 
    $T_{\rm{C}}$ corresponding to arrangements \#1 and \#4 shifted towards higher temperatures than that corresponding to an electric field strength of zero, indicating that controlling magnetism via an electric field is feasible in CaBaCo$_4$O$_7$. 
    Conversely, the $T_{\rm{C}}$ corresponding to arrangements \#2 and \#3 shifted toward lower temperatures. 
    The difference in $T_{\rm{C}}$ between arrangements \#1 (\#4) and \#2 (\#3) is approximately 1 K\@. 
    This behavior indicates that the ferrimagnetic phase was stabilized or destabilized depending on the direction of the applied electric field. 
    Additionally, because no shift in $T_{\rm{C}}$ is observed when comparing arrangement \#1 with \#4 (or \#2 with \#3), it is clear that the sign reversal of $\boldsymbol{P}\times\boldsymbol{M}$ does not contribute to the shift in $T_{\rm{C}}$.

    \begin{figure}
        \centering
        \includegraphics[height=70mm]{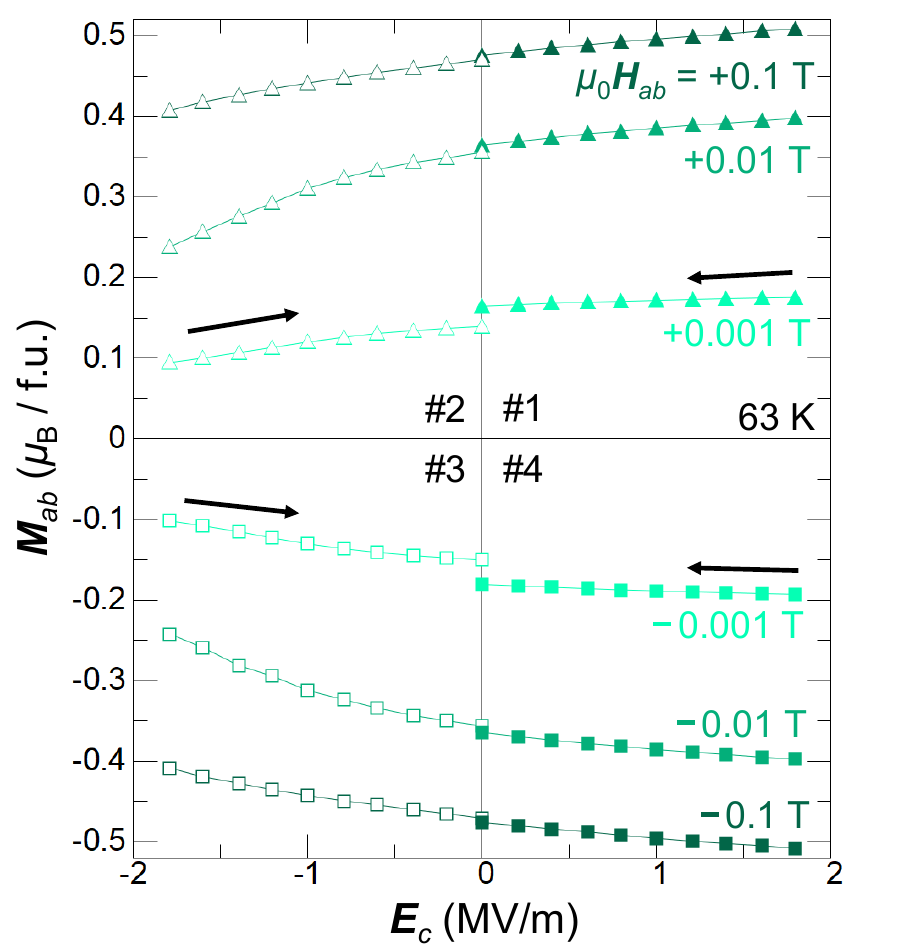}
        \caption{
        (Color online) Electric field dependence of magnetization in magnetic fields with intensities of 0.001, 0.01, and 0.1 T at 63 K in measurement arrangements \#1 to \#4. 
        The solid and open symbols correspond to the positive and negative poling electric fields, respectively, whereas the triangular and square symbols correspond to the positive and negative poling magnetic fields, respectively.
        }
        \label{ME}
    \end{figure}

    To investigate the behavior of magnetization depending on the directions of the applied external fields in more detail, we conducted measurements of the electric-field-induced magnetization in magnetic fields ($\boldsymbol{M}-\boldsymbol{E}$ in $\boldsymbol{H}$) and magnetic-field-induced magnetization in electric fields ($\boldsymbol{M}-\boldsymbol{H}$ in $\boldsymbol{E}$) at a fixed temperature. 
    The result of the $\boldsymbol{M}-\boldsymbol{E}$ in $\boldsymbol{H}$ measurements is shown in Fig.~\ref{ME}\@. 
    The $\boldsymbol{M}-\boldsymbol{E}$ curves were recorded during scans of decreasing electric field without removing the magnetic field after preparing the sample with the poling procedure described above.
    
    At each magnetic field strength, the magnitude of magnetization at $\boldsymbol{E}_c$ = 0 was found to be similar, regardless of the direction of the applied magnetic and electric fields. 
    At $|\boldsymbol{E}_c|$ = 2 MV/m, a comparison of \#1 with \#2 and \#3 with \#4 revealed that $|\boldsymbol{M}_{ab}|$ in \#1 (\#4) is larger than that in \#2 (\#3). 
    However, a comparison between \#1 and \#4, as well as between \#2 and \#3, shows that $|\boldsymbol{M}_{ab}|$ is nearly equal. 
    These results reflect the stabilization (or destabilization) of the ferrimagnetic phase by the application of an external electric field, which is consistent with the shift in $T_{\rm{C}}$ shown in Fig.~\ref{MT}\@.
    
    \begin{figure}
        \centering
        \includegraphics[height=50mm]{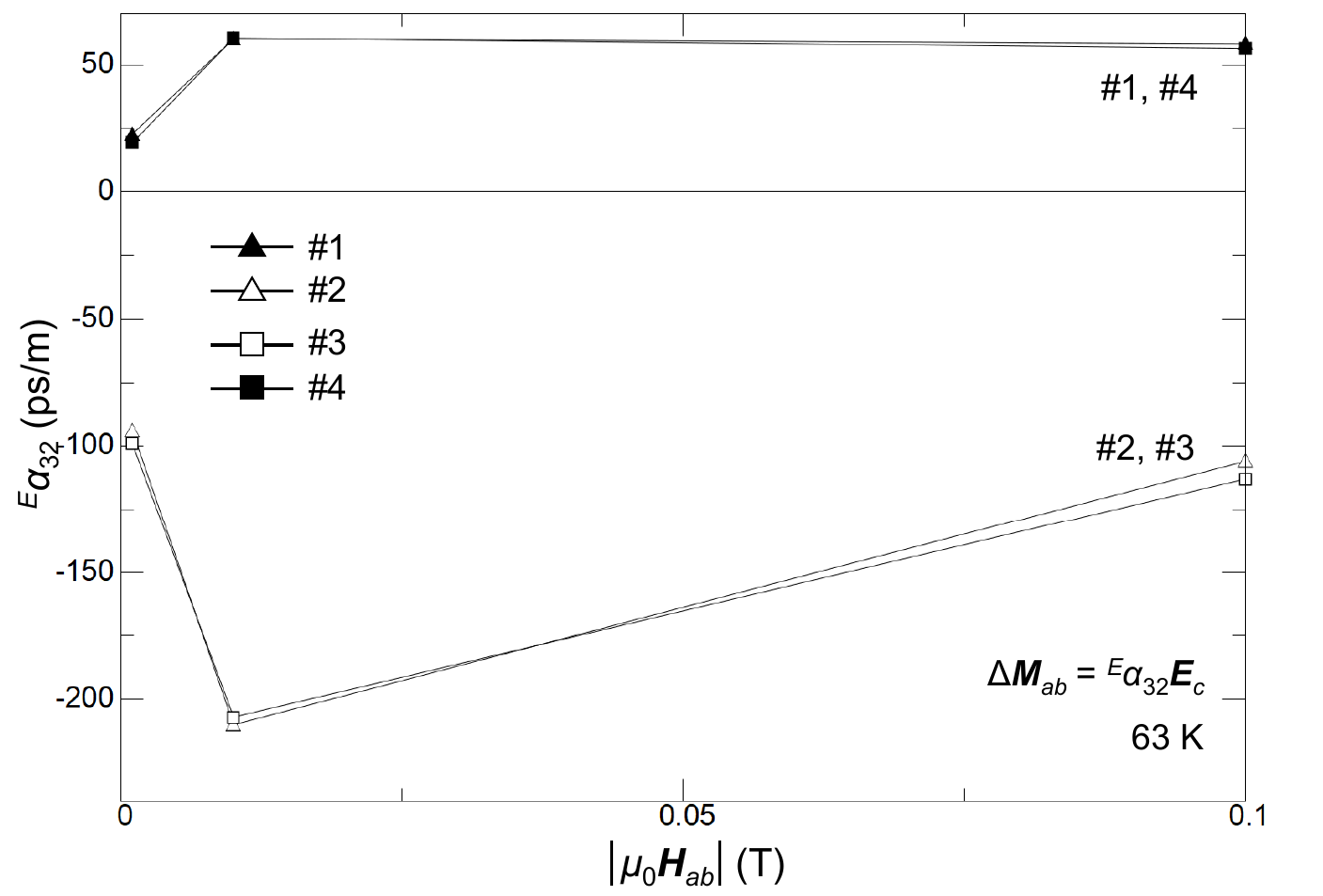}
        \caption{
        Magnetic field dependence of the linear magnetoelectric tensor component, $^E\alpha_{32}$, defined as $\Delta \boldsymbol{M}_{ab}\,=\,^E\alpha_{32}\,\boldsymbol{E}_c$.
        }
        \label{tensor}
    \end{figure}
    
    The ferrimagnetic phase of CaBaCo$_4$O$_7$ belongs to the magnetic point group $m'm2'$, resulting in non-zero off-diagonal components $\alpha_{32} = \alpha_{23}$ of the linear magnetoelectric tensor\cite{Rivera_Euro2009}. 
    Figure~\ref{tensor} shows the absolute magnetic field dependence of the linear magnetoelectric tensor component $^E\alpha_{32}$, defined by the following equation:
    \begin{equation}
        \Delta \boldsymbol{M}_{ab}\,=\,^E\alpha_{32}\,\boldsymbol{E}_c
    \end{equation}
    where $\Delta \boldsymbol{M}_{ab}$ denotes the change in the magnetization from its value at $\boldsymbol{E}_c$ = 0. 
    The value of $^E\alpha_{32}$ was determined by applying a linear least-squares fit to the $\boldsymbol{M}-\boldsymbol{E}$ in $\boldsymbol{H}$ data shown in Fig.~\ref{ME}\@. 
    The sign of $^E\alpha_{32}$ is positive in arrangement \#1 and \#4, where $|\Delta \boldsymbol{M}_{ab}|$ is enhanced by the applied electric field, whereas it is negative in \#2 and \#3, where $|\Delta \boldsymbol{M}_{ab}|$ is suppressed. 
    Furthermore, at each magnetic field strength, the absolute value of $^E\alpha_{32}$ in \#2 and \#3 is larger than that in \#1 and \#4. 
    The magnitude of $|^E\alpha_{32}|$ reached its maximum at $|\mu_0\boldsymbol{H}_{ab}|$ = 0.01 T in all four arrangements. 
    The linear magnetoelectric tensor $^H\alpha_{32}$, as reported in a previous study from the magnetic field dependence of the change in spontaneous electric polarization, is approximately 300 ps/m at 63 K\cite{Caignaert_PRB2013}. 
    This value is on the same order of magnitude as the maximum value of $^E\alpha_{32}$ obtained in this study for arrangements \#2 and \#3. 
    Furthermore, the magnetic-field-induced maximum behavior of $^E\alpha_{32}$ is considered to result from the shift in $T_{\rm{C}}$ caused by the applied magnetic field\cite{Iwamoto_JPC2012}.

    \begin{figure}
        \centering
        \includegraphics[height=50mm]{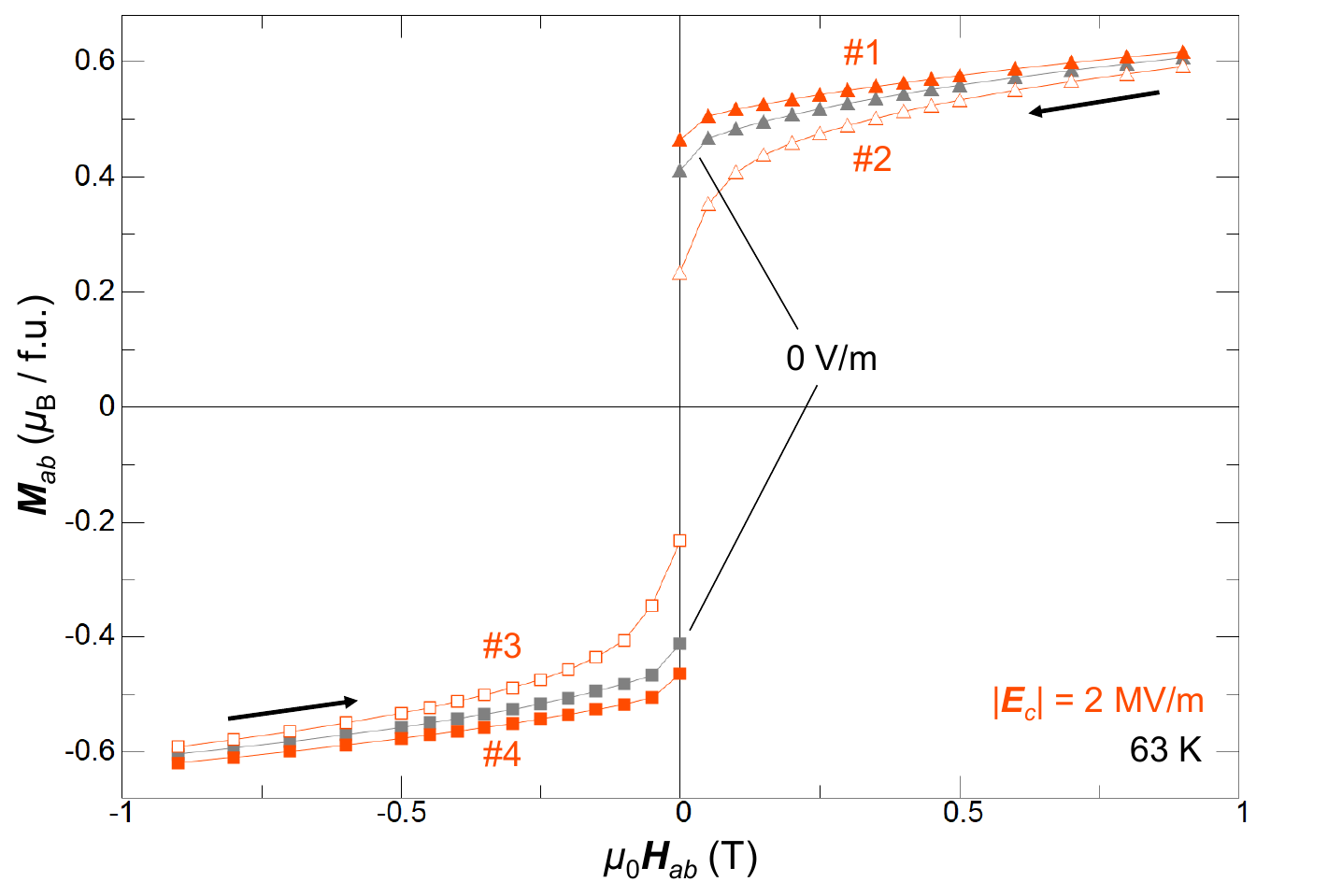}
        \caption{
        (Color online) Magnetic field dependence of magnetization in electric fields with strengths of $\pm$2 MV/m at 63 K. 
        The measurement arrangement corresponding to each plot is as described in the caption of Fig.~\ref{MT}\@.
        }
        \label{MH}
    \end{figure}

    Figure~\ref{MH} shows the results for the $\boldsymbol{M}-\boldsymbol{H}$ in $\boldsymbol{E}$ measurements. 
    The $\boldsymbol{M}-\boldsymbol{H}$ curves were recorded during scans of decreasing magnetic field without removing the poling electric field after preparing the sample with the poling procedure described above.
    
    At zero electric field, the $\boldsymbol{M}-\boldsymbol{H}$ curve exhibited symmetric behavior upon reversal of the magnetic field. 
    In a magnetic field of $\pm$1.0 T, the magnitude of the magnetization in all four arrangements matches the value at zero electric field. 
    As the applied magnetic field increased, $T_{\rm{C}}$ shifted to a higher temperature\cite{Iwamoto_JPC2012}. 
    In an absolute magnetic field of 1.0 T, spin fluctuations at 63 K were reduced, leading to a weakened effect of the electric field on magnetization. 
    However, the magnitude of the remanent absolute magnetization increased in \#1 and \#4, but decreased in \#2 and \#3 compared with the zero electric field. 
    This behavior is consistent with the previously described magnetization trends and indicates the stabilization (or destabilization) of the ferrimagnetic phase by an external electric field. 
    In the \#2(\#3) arrangements, the decrease in $|\boldsymbol{M}_{ab}|$ from its zero electric field value corresponds to a temperature increase of approximately 0.5 K, as estimated from Fig.~\ref{MT}\@. 
    At 63 K, CaBaCo$_4$O$_7$ is a sufficiently good insulator, and the Joule heating caused by leakage current under an applied electric field of 2 MV/m is negligible. 
    Therefore, the electrically induced change in magnetization observed in this study can be considered an intrinsic phenomenon.

    \begin{figure}
        \centering
        \includegraphics[height=50mm]{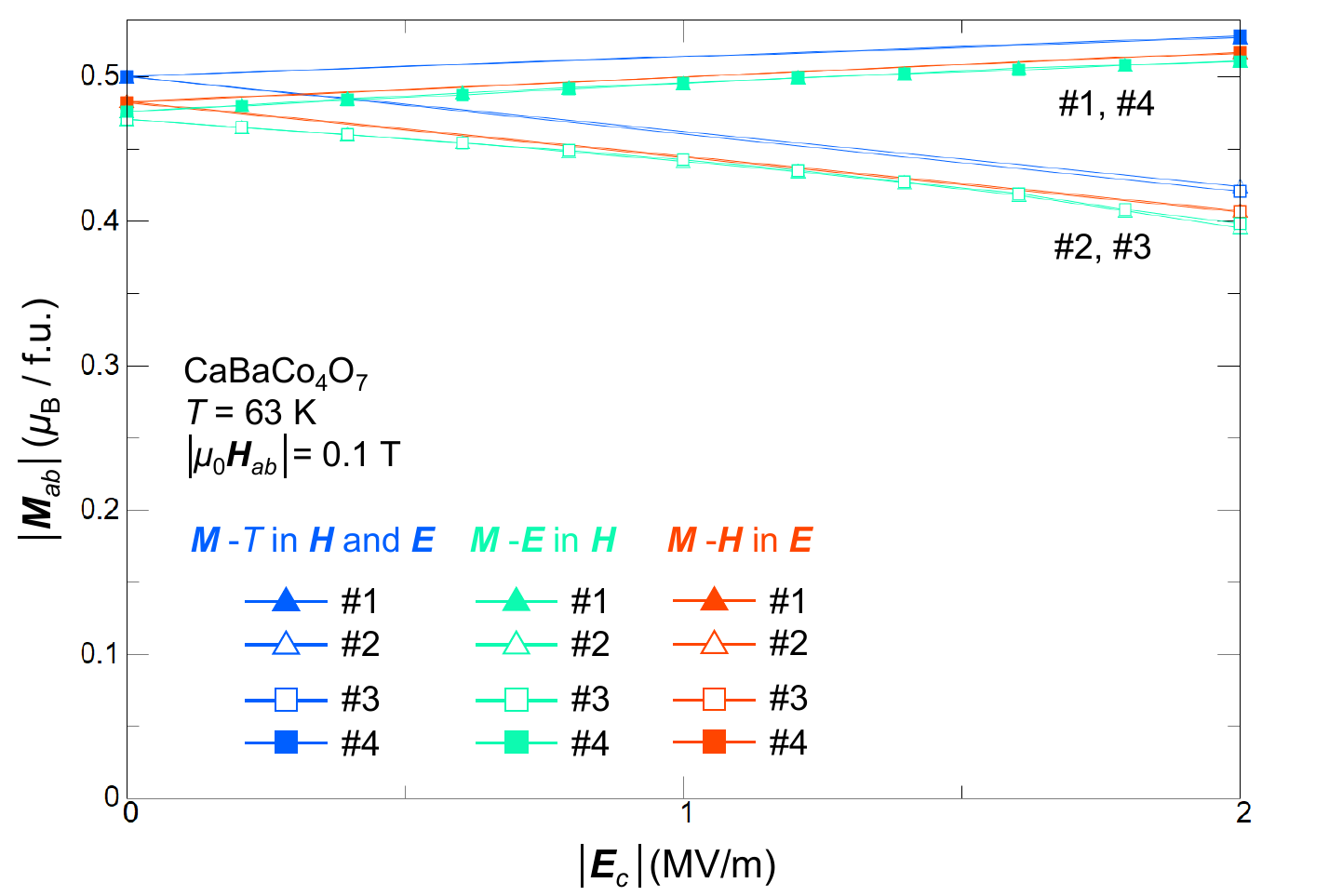}
        \caption{
        (Color online) Electric field dependence of magnetization in $|\mu_0\boldsymbol{H}_{ab}|$ = 0.1 T from the three measurements shown in Figs.~\ref{MT}, \ref{ME}, and \ref{MH}\@. 
        Note that the electric field and magnetization are plotted as their positive absolute values.
        }
        \label{summary}
    \end{figure}

    The results obtained from the three magnetization measurements, i.e., $\boldsymbol{M}-T$ in $\boldsymbol{H}$ and $\boldsymbol{E}$, $\boldsymbol{M}-\boldsymbol{E}$ in $\boldsymbol{H}$, and $\boldsymbol{M}-\boldsymbol{H}$ in $\boldsymbol{E}$, are summarized in Fig.~\ref{summary}\@. 
    In all three measurements, the absolute values of magnetization, $|\boldsymbol{M}_{ab}|$, increased for arrangements \#1 and \#4, while they decreased for \#2 and \#3 with an increase in the absolute value of the electric field $|E_c|$. 
    In addition, the absolute values of the magnetization at $|\boldsymbol{E}_c|$ = 2 MV/m for the three measurements were almost equal for each set of arrangements. 
    These results clearly demonstrate the electric-field-direction dependence of the static ME effect in the pyroelectric CaBaCo$_4$O$_7$ crystal.
    
    This behavior can be phenomenologically understood by assuming a fixed direction for the change of spontaneous electric polarization $\Delta \boldsymbol{P}$, as depicted in Fig.~\ref{CS}(b)\@. 
    The free energy for the ferrimagnetic (FiM) and antiferromagnetic (AFM) phases are given by $F_{\rm FiM} = - P_{\rm FiM}E_c - M_{\rm AFM}B_{ab} - \,^E\alpha_{32}E_cB_{ab}$ and $F_{\rm AFM} = - P_{\rm AFM}E_c - M_{\rm AFM}B_{ab}$, respectively.
    Here, for simplicity, we denote the external magnetic field by $B_{ab}$ ($= \mu_0 H_{ab}$). 
    Note that the magnetoelectric tensor is zero in the AFM phase\cite{Caignaert_PRB2013}.
    Therefore, the change in free energy at the AFM to FiM phase transition, $\Delta F = F_{\rm FiM} - F_{\rm AFM}$, can be expressed as:
    \begin{equation}
        \Delta F\,=\, - \Delta P E_c - \Delta M B_{ab} - \,^E\alpha_{32}\,E_cB_{ab}
    \end{equation}
    where $\Delta P = P_{\rm FiM} - P_{\rm AFM}$ and $\Delta M =  M_{\rm FiM} - M_{\rm AFM}$\@.
    The term of magnetoelectric coupling, $\,^E\alpha_{32}\,E_cB_{ab}$, is negligibly small compared to the terms of $\Delta P E_c$ and $\Delta M B_{ab}$\@.
    Furthermore, $\Delta M B_{ab}$ is invariant under the reversal of the magnetic field.
    Therefore, the free energy change $\Delta F$ for each measurement arrangements is dominated by the $\Delta P E_c$ term. 
    Consequently, the ferrimagnetic phase is stabilized when the $\Delta P$ and $E_c$ are parallel, resulting in  $\Delta P E_c > 0$.
    Conversely, the ferrimagnetic phase is destabilized when they are antiparallel.
    Similar control of magnetic phase stability by the electric field direction has also been reported in the skyrmion host material Cu$_2$OSeO$_3$\cite{Okamura_NatCommun2016}.
    To further investigate the electric-field-direction dependence of the ME effect in CaBaCo$_4$O$_7$, the remaining twin structure should be detwinned in the \textit{ab} plane, and measurements can be recorded to distinguish between the \textit{a} and \textit{b} axes\cite{Arai_JPSC2022}. 
    
    The magnetization in other polar materials may also be controlled in a direction-dependent manner by considering the crystal orientation, as demonstrated in our experiments, thereby enabling a significant enhancement of the ME effect. 
    In addition, the dependence of the $T_{\rm{C}}$ shift on the direction of the electric field suggests the possibility of inducing and erasing spontaneous magnetization using an electric field\cite{Shimamura_APL2012}. 
    Although the observed effect remains small and the operating temperature is currently limited, this experiment demonstrates a new pathway for controlling magnetic properties. 
    The ability to stabilize or destabilize a magnetic phase simply by changing the direction of an external electric field, as shown in this study, suggests the potential for pyroelectric-magnetic materials to be utilized in novel spintronics devices.
    
\acknowledgment
    We would like to thank H. Endoh for his help in growing single crystals. 
    This work was partly supported by JSPS KAKENHI Grant Numbers JP15K05184, JP16K05420, and JP19K03745.

\end{document}